\documentclass[aps,pra,twocolumn,superscriptaddress]{revtex4-2}
\usepackage[utf8]{inputenc}
\usepackage{soul}
\usepackage{textcomp}
\usepackage[english]{babel}
\usepackage{graphicx}
\usepackage{dcolumn}
\usepackage{bm}
\usepackage{amsmath}
\usepackage{verbatim}     
\usepackage[dvipsnames]{xcolor}       
\usepackage[normalem]{ulem}   
\usepackage{bbold}
\usepackage{mathrsfs}
\usepackage[mathscr]{eucal}
\usepackage{physics}
\usepackage{algorithmic}
\usepackage{gensymb}
\usepackage[bottom]{footmisc} 

\usepackage{hyperref}
\hypersetup{
   pdfpagemode=None,
   pdfstartpage=1,
   pdfmenubar=true,
   pdftoolbar=true,
   colorlinks = true,
   linkcolor=blue,
   citecolor=blue,
   urlcolor=blue,
   bookmarksopen=false
 } 

 \usepackage{booktabs}
 \usepackage{float}

\newcommand{\qo}[1]{``#1''}   

\begin{document}

\title{Predicting atmospheric turbulence for secure quantum communications in free space}

\author{Tareq Jaouni}
\thanks{These authors contributed equally to this work.}
\affiliation{Nexus for Quantum Technologies, University of Ottawa, Ottawa ON, Canada, K1N 5N6}

\author{Lukas Scarfe}
\thanks{These authors contributed equally to this work.}
\affiliation{Nexus for Quantum Technologies, University of Ottawa, Ottawa ON, Canada, K1N 5N6}

\author{Frédéric Bouchard}
\affiliation{National Research Council of Canada, 100 Sussex Drive, Ottawa ON, Canada, K1A 0R6}

\author{Mario Krenn}
\affiliation{Max Planck Institute for the Science of Light, Staudtstrasse 2, 91058 Erlangen, Germany}

\author{Khabat Heshami}
\affiliation{National Research Council of Canada, 100 Sussex Drive, Ottawa ON, Canada, K1A 0R6}
\affiliation{Nexus for Quantum Technologies, University of Ottawa, Ottawa ON, Canada, K1N 5N6}

\author{Francesco Di Colandrea}
\email{francesco.dicolandrea@unina.it}
\affiliation{Nexus for Quantum Technologies, University of Ottawa, Ottawa ON, Canada, K1N 5N6}
\affiliation{Dipartimento di Fisica, Universit\`{a} degli Studi di Napoli Federico II, Complesso Universitario di Monte Sant'Angelo, Via Cintia, 80126 Napoli, Italy}

\author{Ebrahim Karimi}
\affiliation{Nexus for Quantum Technologies, University of Ottawa, Ottawa ON, Canada, K1N 5N6}
\affiliation{National Research Council of Canada, 100 Sussex Drive, Ottawa ON, Canada, K1A 0R6}
\affiliation{Max Planck Institute for the Science of Light, Staudtstrasse 2, 91058 Erlangen, Germany}

\begin{abstract}
Atmospheric turbulence is the main barrier to large-scale free-space quantum communication networks. Aberrations distort optical information carriers, thus limiting or preventing the possibility of establishing a secure link between two parties. For this reason, forecasting the turbulence strength within an optical channel is highly desirable, as it allows for knowing the optimal timing to establish a secure link in advance. Here, we train a Recurrent Neural Network, TAROCCO, to predict the turbulence strength within a free-space channel. The training is based on weather and turbulence data collected over 9 months for a 5.4~km intra-city free-space link across the City of Ottawa. The implications of accurate predictions from our network are demonstrated in a simulated high-dimensional Quantum Key Distribution protocol based on orbital angular momentum states of light across different turbulence regimes. TAROCCO will be crucial in validating a free-space channel to optimally route the key exchange for secure communications in real experimental scenarios.
\end{abstract}

\maketitle

\section{Introduction}
The ability to forecast weather conditions is a relevant achievement in modern society, impacting agriculture~\cite{agronomy13030625}, energy management~\cite{LAZOS2014587}, climate science~\cite{kendon2014heavier}, and public health~\cite{hazard}. Traditional forecasts involve tangible parameters such as temperature, wind speed, and precipitation, while the complex phenomenon of atmospheric turbulence has mainly been restricted to astronomical observations~\cite{roddier1981v}, and only recently extended to free-space communications~\cite{1025501,PhysRevA.97.012321}. Atmospheric turbulence originates from the rapid variation in time of different weather parameters, which results in optical signals experiencing a continuously varying refractive index when traveling through the atmosphere~\cite{wyngaard1992atmospheric}. This severely affects the reliability of optical experiments realized in \qo{free space}, which has led to the dominance of fibre-based communications networks. Scintillation of low-power light, beam wandering, and wavefront distortion are representative examples of common optical aberrations, globally quantified by the structure parameter $C_{n}^2$~\cite{kolmogorov}.

Atmospheric turbulence significantly limits optical communications, both in the classical~\cite{Anguita:08} and quantum~\cite{Tyler:09} regime. Typical effects are high crosstalk and lower power transmission. For these reasons, adaptive-optics correction systems have been proposed and experimentally demonstrated, showing promising results for free-space Quantum Key Distribution (QKD) in turbulent channels~\cite{pugh2020adaptive,Zhao:20,scarfe2023fast}. These achievements set the baseline for future ground-to-ground~\cite{PhysRevApplied.16.044027} and ground-to-satellite~\cite{PhysRevApplied.16.014067} configurations. In the broader context of quantum networks, predicting the trend of atmospheric turbulence is key to establishing a secure connection between two parties, leading to informed decisions about the future usage of available channels within the network.


Based on historical weather data, standard approaches to forecasting atmospheric turbulence relied upon empirical models~\cite{BINKOWSKI1979247}, remote sensing~\cite{rafalimanana} and, more recently, artificial neural networks~\cite{wang2014estimation,Grose:23,Bi2023,AURORA}. 
In particular, Grose and Watson explored the application of different instances of Recurrent Neural Networks (RNNs) to predict $C_n^2$ values up to a few hours in advance~\cite{Grose:23}. These architectures are specifically designed to model sequential data, which makes them suitable candidates for capturing the temporal dependencies within the input weather data in the form of a time series. Although excellent levels of accuracy have been reported in this first case study, the final implementation suffered from relevant limitations, most importantly the lack of real night-time data, a training dataset only spanning a single season, and a rigid fixed-time prediction in the future~\cite{Grose:23}.

In this paper, we train TAROCCO~\footnote[1]{In Italian, Tarocco means tarot cards, which are used in fortune telling.}, an RNN processing dataset encompassing 9 months of weather data with a minute-by-minute time resolution to forecast $C_n^2$ values within a 5.4~km intra-city channel over the City of Ottawa. Our scheme outputs predictions within a flexible number of hours in the future (up to 12 hours) and with a custom time resolution (down to one minute). This computational toolbox provides the preliminary validation step of a turbulent channel for near-term QKD experiments. The significance of this result is numerically demonstrated by simulating a high-dimensional BB84 QKD protocol employing spatial modes of light under different turbulent regimes. In principle, all free-space experiments utilizing structured light can benefit from this tool, since knowing the turbulence strength in advance can allow for predictions of the success rate of these experiments~\cite{KrennOAM143,Krenn_2014,lavery,Sit:17}. The same analysis can apply to nearby satellite ground stations, which would allow for determining the optimal channel for maximum key exchange on a given satellite pass.

\section{Theory}
\subsection{Atmospheric turbulence}
\label{sec:theory}

Atmospheric turbulence can be quantified in terms of the structure parameter of the refractive index $C_n^2$. The $C_n^2$ is defined as the variance of the refractive index over a given optical path, normalized to the path length~\cite{tatarski1961wave}:
\begin{equation}
    C_n^2 = \overline{\frac{(n(\vec{x})-n(\vec{x}+\vec{r}))^2}{r^{\frac{2}{3}}}} \approx \frac{\text{var}( n)}{r^{\frac{2}{3}}}, 
\end{equation}
where $n,\vec{x}, \text{and } \vec{r}$ are the refractive index of the atmosphere, the position along the path, and the distance from the sender to the receiver, respectively, with the average taken over all positions within the path. 

The effect of atmospheric turbulence on optical beams can be modeled as a random sequence of phase masks dislocated along the path. By randomizing the phase screens in multiple realizations, it is possible to retrieve the average effect of the turbulence on the input beam with good approximation ~\cite{doi:10.10520/EJC96781}. Zernike modes, which are traditionally employed to model optical aberrations~\cite{BornWolf}, provide a natural choice to generate the phase masks. In particular, the value of $C_n^2$ can be directly related to the variance of each Zernike mode~\cite{Noll:76}. 

To measure the value of $C_n^2$ in our channel, we employ scintillometry~\cite{Ward_2017}. By continuously measuring the optical intensity of a beam propagating from the sender to the receiver, relevant statistical parameters can be extracted to retrieve the structure parameter over a given time period.

\subsection{Quantum Key Distribution}
Quantum key distribution (establishment) is a technique introduced by Bennett and Brassard with the well-known BB84 protocol~\cite{bennett2014quantum}, which utilizes the principles of quantum mechanics to enable provably secure communications between two parties, by establishing a shared secret key. Security is guaranteed when the error in transmission and measurement of the quantum information is below a given threshold, depending on the dimension of the protocol~\cite{Gottesman:security}. Extending QKD to high-dimensional states allows for higher information capacity (typically expressed in \qo{bits per photon}), higher tolerance to errors, and innovative protocols~\cite{QKDhighdim}.

In a $d$-dimensional error-free QKD protocol, the number of informational bits per photon is $R(d)=\log_2(d)$. Turbulence-induced errors within a free-space channel result in a diminished key rate. For a high-dimensional BB84 protocol~\cite{Sheridan:quditsecure}:
\begin{equation}
    R(d,e_q) = \log_2(d) -2 h(e_q),
    \label{eq:keyrate}
\end{equation} 
where ${h(x)= -x\log_2(x/(d-1))-(1-x)\log_2(1-x)}$ is the Shannon entropy, and $e_q$ is the error rate. When $R(d,e_q)<0$, security is not guaranteed. 

It must be noted that QKD protocols are typically not used to send messages, but only to establish a shared key through which classical communications can be securely encrypted. These keys can also be generated and stored for later use, even if the quantum channel is incapable of generating new keys~\cite{Sharma:storekey}. 

Further reading on the practical implementations of QKD can be found in Ref.~\cite{9374032}.

\begin{figure*}[!htb]
	\begin{center}
		\includegraphics[width=2.0\columnwidth]{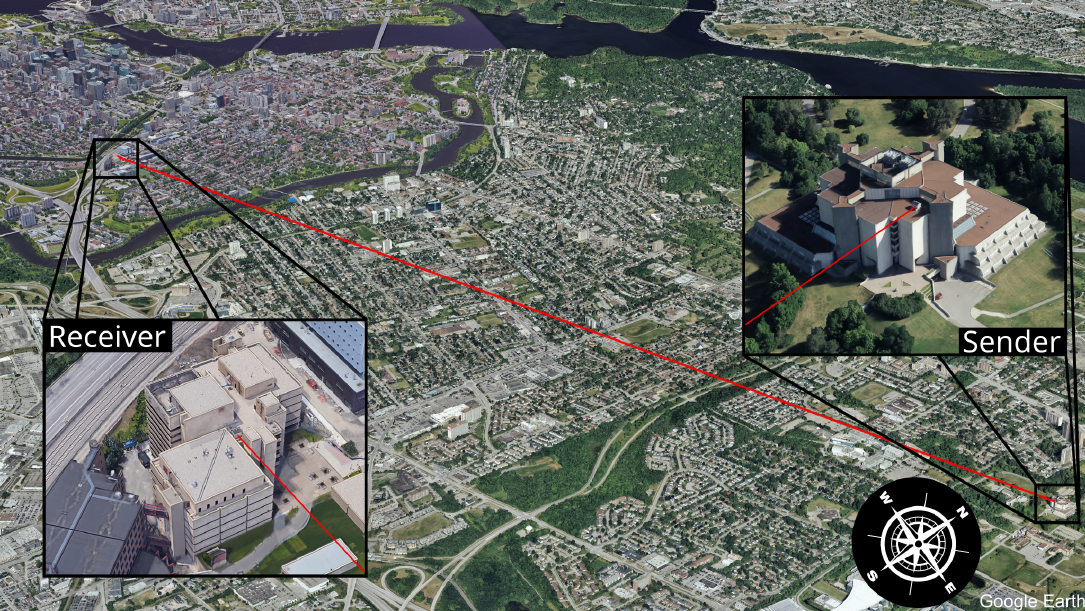}
		\caption{\textbf{Channel path.} The path over which the turbulence data was collected from July 2023 to March 2024. The receiver is on the University of Ottawa campus, while the sender is on the Canadian National Research Council 5.4~km ENE. This configuration was chosen to minimize the background levels caused by the sunlight. 
  }
		\label{fig:Google Maps}
	\end{center}
\end{figure*}

\section{Methodology}
\subsection{Data Collection and Preparation}
We employ the Scintech BLS450 Large Aperture Scintillometer to record the value of $C_n^2$ over our channel~\cite{Scintech}. The system has been active 24/7 measuring the turbulence of the channel for 9 months. An aerial view of the monitored channel is provided in Fig.~\ref{fig:Google Maps}. The $C_n^2$ values have been recorded once per minute. 
Fog and heavy snowfall can cause system outages, corresponding to missing data points in our training set. To remove spurious information deriving from high-frequency noise, a moving average is performed for each data point over a one-hour time window. 

Other weather parameters, such as temperature, solar radiation, and humidity, have been obtained from Environment and Climate Change Canada ~\cite{Meteor}. The weather station at which the parameters are measured lies 5 km SW from the receiver. For these parameters, no additional processing or filtering was applied. The averaged $C_n^2$ values are time correlated with the other meteorological data, and finally separated into batches for training the RNN. 

The input layer of the network consists of 12 hours of minute-by-minute weather data, with $n$ hours of future $C_n^2$ values as the target output. The dataset is divided into a training part where the network learns, a validation dataset to provide feedback on the network's training, and a test dataset to evaluate the network's performance on unseen data. All input and output features are normalized according to: 
\begin{equation}
    X \rightarrow \frac{X - \text{min}(X_{T})}{\text{max}(X_{T}) - \text{min}(X_{T})},
\label{eqn:norm}
\end{equation}
\noindent where $\text{min}(X_{T})$ and $\text{max}(X_{T})$ denote, respectively, the minimum and maximum of the feature $X$ over the training dataset. 

\subsection{Recurrent Neural Network}

\label{sec:rnn}

\begin{figure*}[!htb]
	\begin{center}
		\includegraphics[width=0.9\linewidth]{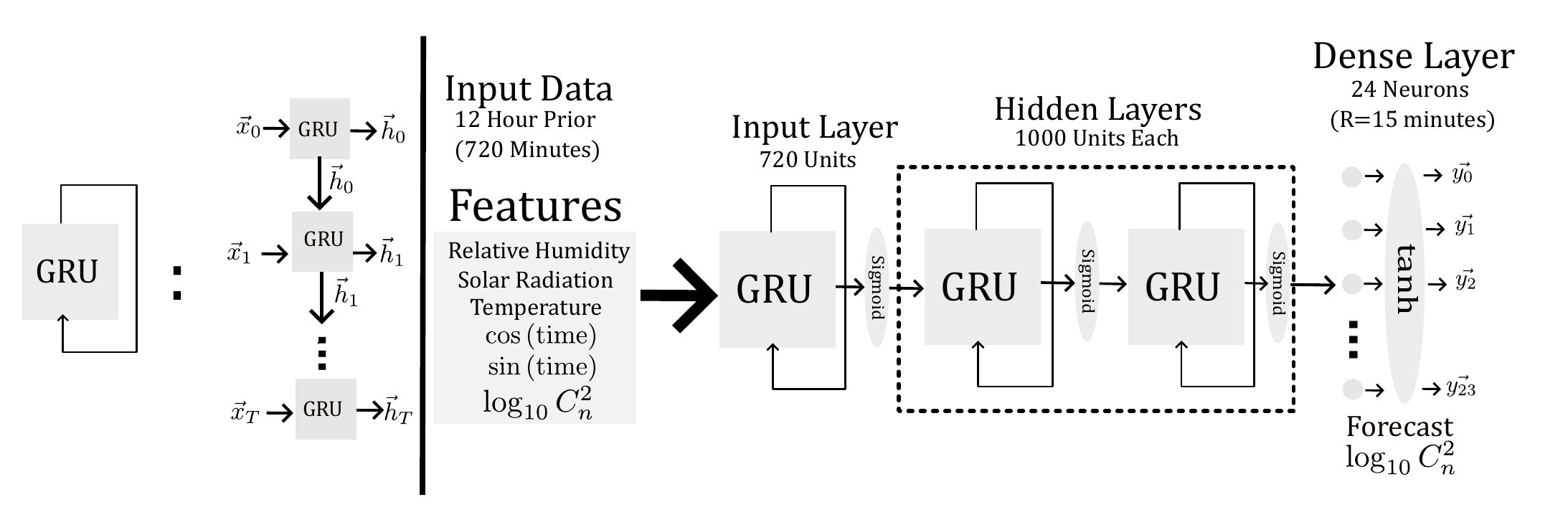}
		\caption{\textbf{Neural network architecture.} The network processes 12 hours of input data. The data is fed through a series of Gated Recurrent Unit (GRU) layers, allowing the network to learn long-term trends through time in the input series. The hidden state variable $\vec{h}$ is standardized using nonlinear activation layers at each step. The network outputs the forecast in a single shot, through the final fully connected dense layer.}
		\label{fig:RNN}
	\end{center}
\end{figure*}

We train a Gated Recurrent Unit (GRU) neural network, named TAROCCO, to forecast the evolution of the $C_{n}^{2}$ values over time. GRUs are a simplified derivative of the long short-term memory (LSTM) recurrent units, which capture long-term features of the data. Figure~\ref{fig:RNN} illustrates our neural network architecture, which takes 12 hours of prior combined meteorological and scintillometer data as input and outputs the $C_{n}^{2}$ forecast over 6 hours. 
Following the Permutation Feature Importance method detailed in Appendix~\ref{appendix: grid}, temperature (\degree C), solar radiation (kJ/m$^{2}$), relative humidity (\%), $\log_{10}{C_{n}^{2}}$, and the UTC time (s) are used as input features. The time feature $t$ is made periodic over one day with the following map: 
\begin{equation}
\begin{split}
    t_x &= \cos{\left(\frac{2\pi t}{T}\right)},\\
    t_y &= \sin{\left(\frac{2\pi t}{T}\right)},
\end{split}
\end{equation}
where $T=86400$ s. 

Our network outputs the turbulence forecast through a fully connected dense layer that takes as input the final hidden state of the incident GRU layer. The number of output neurons is given by ${N_\text{out}=H/R}$, where $H$ is the desired number of hours in the future (6 hours) and $R$ is the desired output time resolution (15 minutes). 

The cost function used for training is the Mean Squared Error (MSE): \begin{equation}
    \text{MSE} = \frac{1}{N}\sum_{i=1}^{N} |\vec{y}_{i, \text{pred}} - \vec{y}_{i, \text{true}}|^{2}, 
\end{equation}
where $\vec{y}_{\text{pred}}$ and $\vec{y}_{\text{true}}$ denote the predicted and true output $C_{n}^{2}$ forecast, and $N$ is the size of the evaluated dataset. The complete list of hyperparameters used for training is reported in Appendix~\ref{appendix: train-hype}.

\section{Results}
\subsection{$C_n^{2}$ forecast} 
\label{sec:results}
The network is validated on two test datasets, one spanning October 2023 and the other from February 14th to March 14th, 2024. For these tests, we employed two separate architectures, respectively trained on a set deprived of the corresponding test set. The dataset from October 2023 is specifically chosen as it features a significantly variable evolution of the $C_{n}^{2}$ compared to other months, which makes it a challenging validation. The performance of the network is evaluated as the average root mean squared error, defined as ${\overline{\text{RMSE}} = \frac{1}{N}\sqrt{\text{MSE}}}$. This error is computed using the predicted and actual ${\log_{10}{C_{n}^{2}}}$ values, normalized according to Eq.~\eqref{eqn:norm}. In the following, we report the results for 6 hours in advance with a 15-minute time resolution. However, we have observed adequate results for up to 12 hours in advance, even with a time resolution down to 1 minute.  

\begin{figure*}[!htb]
	\begin{center}
    \includegraphics[width=1\linewidth]{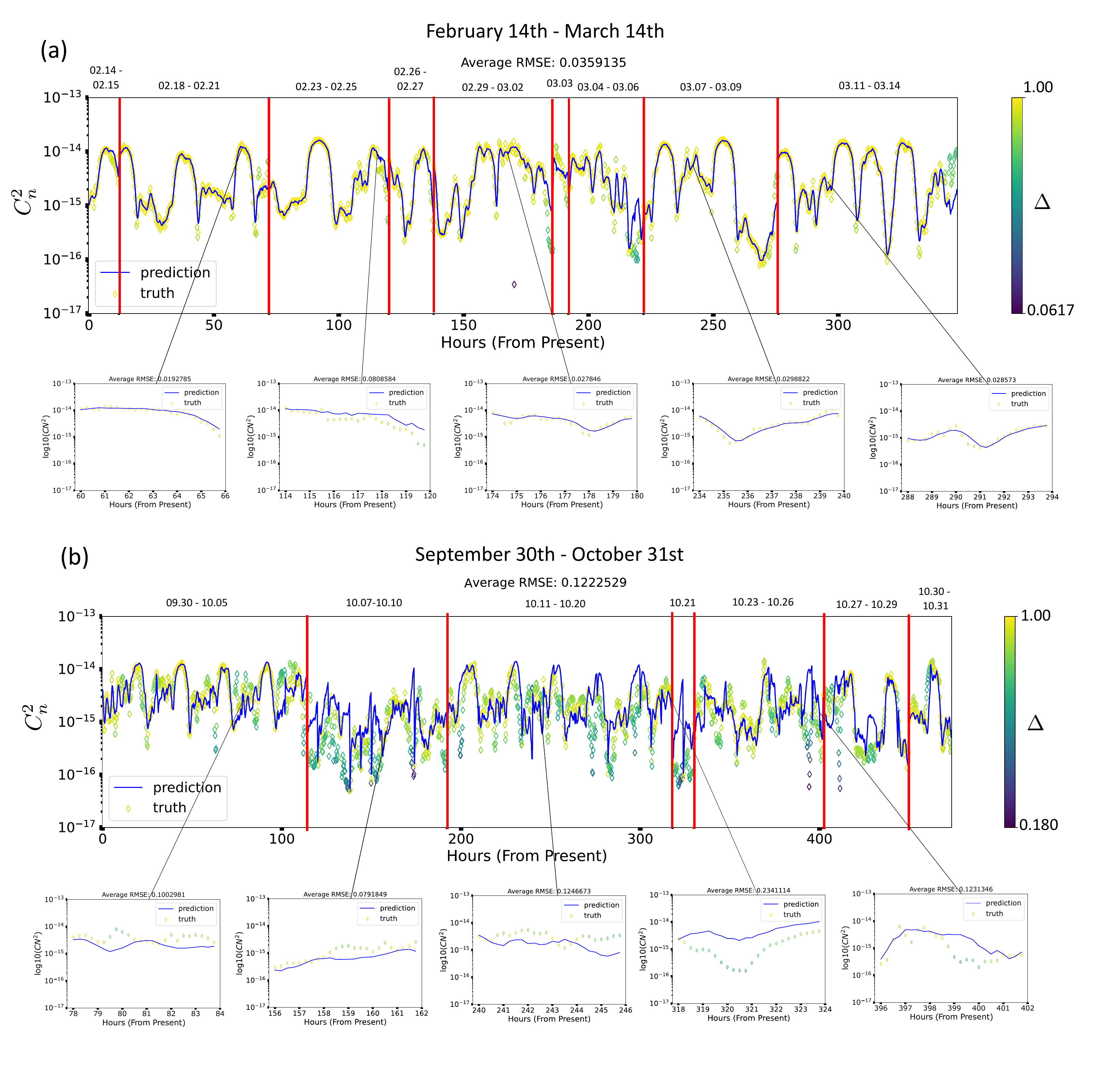}
		\caption{\textbf{Predictions on the test sets.} Network predictions on the months (a)~from mid-February to mid-March 2024 and (b)~of October. We have our trained model forecast 6 hours in the future, then cascade individual predictions to cover one month. Red lines are missing data from the scintillometer system. We also indicate the dates between these discontinuities in the dataset. A one-hour moving average is applied to the training set to remove high-frequency noise. The scatter points refer to the raw data (before the moving average is performed). The insets show examples of individual 6-hour predictions, equally spaced throughout the month. Excellent performances are observed on average, with larger deviations occurring in correspondence with strong feature variations.}
		\label{fig:Results}
	\end{center}
\end{figure*}

Figure~\ref{fig:Results}(a) and (b) plot the prediction of  $C_{n}^{2}$ values over the above-mentioned periods. The plots are obtained by cascading 6-hour predictions sequentially in time. The colorscale encodes the accuracy on the final prediction: ${\Delta(\text{RMSE}) = 1-\text{RMSE}}$, with higher values of $\Delta$ indicating better performances. A typical day features $C_{n}^{2}$ peaks at $10^{-14}$ during the afternoon, dropping in the evening to values approaching $10^{-16}$, followed by a local peak overnight around $10^{-14}$. The insets of both panels show representative examples of individual 6-hour predictions, generated from portions of the test set equally spaced in time. In both cases, the network performs very well in matching the actual $C_{n}^{2}$, with an average RMSE within the order of $10^{-2}$ $(\Delta \geq 0.90)$. It also shows a certain degree of robustness to minor deviations from the trivial periodic trend. However, stronger variations in the feature trends represent a greater challenge for the network, in some cases with deviations within the order of $10^{-1}$ ($\Delta < 0.90$).

\subsection{Numerical QKD experiment} 
\label{sec:qkd}
We numerically investigate the performance of an 8-dimensional BB84 protocol, by simulating a realistic key exchange across the channel with increasing turbulence strengths. In particular, weak (${C_{n}^2=10^{-16}}$), moderate (${C_{n}^2=10^{-15}}$), and strong (${C_{n}^2=10^{-14}}$) turbulence are investigated. The high-dimensional protocol exploits Laguerre-Gaussian (LG) spatial modes of light, carrying a discrete amount of orbital angular momentum (OAM) $\ell \hbar$~\cite{PhysRevA.45.8185}, encoding a qu$d$it of information. The Fourier-conjugate basis (ANGLE modes) is used as a Mutually Unbiased Basis (MUB). The explicit expression of these modes in the position representation is provided in Appendix~\ref{app:modes}. 

\begin{figure*}[!htb]
	\begin{center}
		\includegraphics[width=\linewidth]{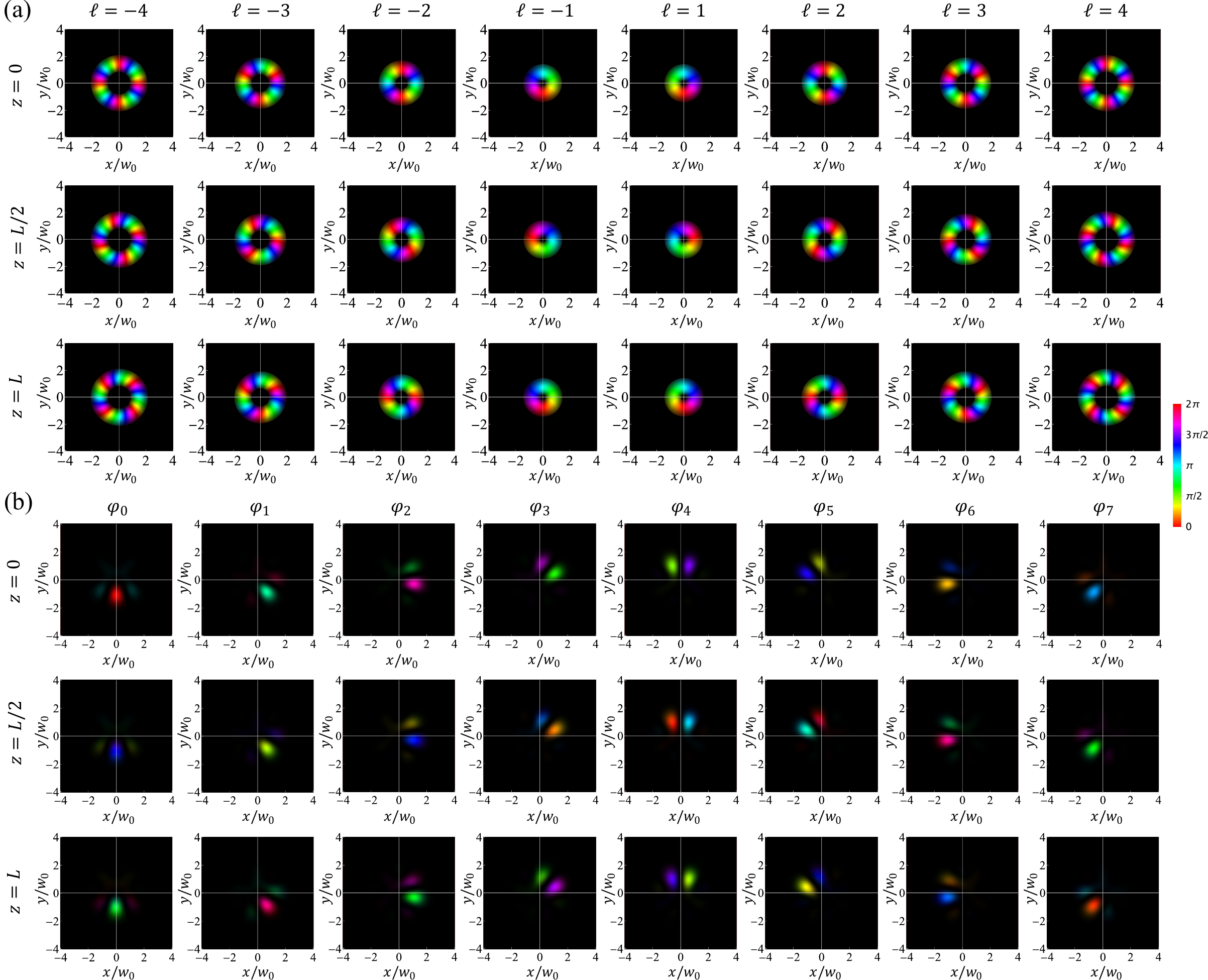}
		\caption{\textbf{8-dimensional QKD.} (a) 8-dimensional OAM space, spanning modes from ${\ell=-4}$ to ${\ell=4}$, excluding ${\ell=0}$. (b)~ANGLE modes $\lbrace{\varphi_0,...,\varphi_7} \rbrace$ provide a possible MUB for the high-dimensional protocol. The input field at ${z=0}$, and the resulting field upon propagation at ${z=L/2}$ and ${z=L}$, with $L=5.4$~km being the length of the free-space link, are provided for each mode. Opacity and hue colorscale encode the amplitude and the phase of the field, respectively.}
		\label{fig:modes_noaberr}
	\end{center}
\end{figure*}

The parameters used in our simulated experiments are: ${w_0=8~\text{cm}}$, ${\lambda=810~\text{nm}}$, ${L=5.4~\text{km}}$, ${D=30~\text{cm}}$, where $w_0$ is the beam waist at ${z=0}$, $\lambda$ is the operating wavelength, $L$ is the channel length, and $D$ is the receiver aperture. Figure~\ref{fig:modes_noaberr}(a) and (b) show the OAM and ANGLE modes in the sender plane (${z=0}$), halfway across the channel (${z=L/2}$), and at the receiver end (${z=L}$), with the opacity and the hue colorscale encoding the amplitude and the phase of the field, respectively. The ${\ell=0}$ mode is removed from the OAM basis, due to non-negligible crosstalks with adjacent modes~\cite{scarfe2023fast}. In our simulations, the effect of turbulence is modeled as a single phase mask located at ${z=0}$, which is generated from a random combination of Zernike modes, where the contribution of each mode is extracted within the corresponding variance associated with the simulated $C_{n}^2$ \cite{Noll:76}. However, a more general approach would require including multiple phase objects across the beam propagation, as discussed in Sec.~\ref{sec:theory}. For each level of turbulence, we run 100 numerical experiments, from which the average crosstalk matrices are extracted. 

\begin{figure*}[!htb]
	\begin{center}
		\includegraphics[width=\linewidth]{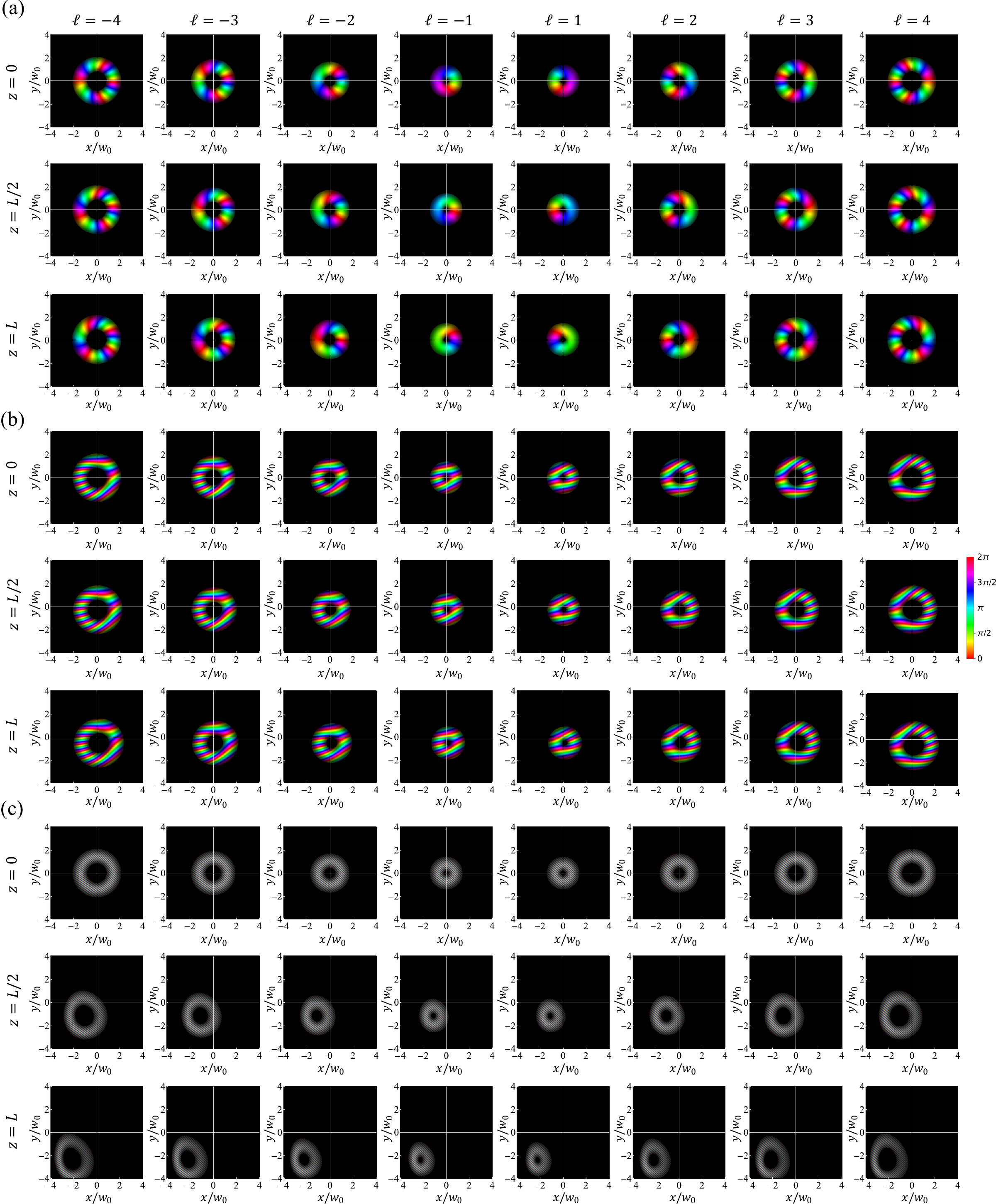}
		\caption{\textbf{Aberrated OAM modes.} Aberrations affect the OAM beam propagation over the channel. Different regimes are considered: (a)~weak ($C_n^2=10^{-16}$), (b)~moderate ($C_n^2=10^{-15}$), and (c)~strong turbulence ($C_n^2=10^{-14}$). The panels refer to the realization yielding the least secure communication, i.e., the one minimizing the diagonal overlap integrals.}
		\label{fig:modes_oam_turb}
	\end{center}
\end{figure*}

\begin{figure*}[!htb]
	\begin{center}
		\includegraphics[width=\linewidth]{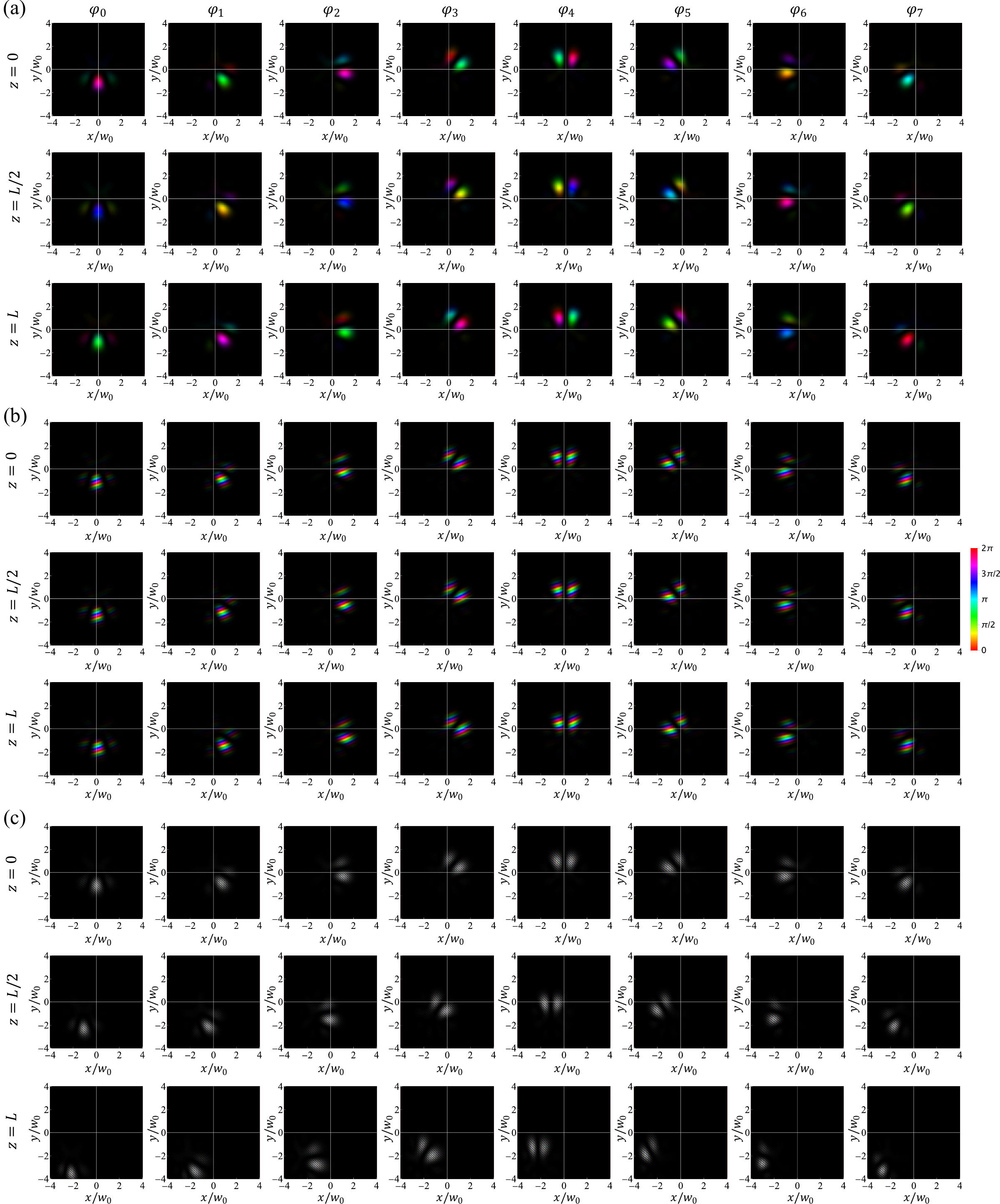}
		\caption{\textbf{Aberrated ANGLE modes.} Aberrations affect the ANGLE beam propagation over the channel. Different regimes are considered: (a)~weak ($C_n^2=10^{-16}$), (b)~moderate ($C_n^2=10^{-15}$), and (c)~strong turbulence ($C_n^2=10^{-14}$). The panels refer to the realization yielding the least secure communication, i.e., the one minimizing the diagonal overlap integrals.}
		\label{fig:modes_angle_turb}
	\end{center}
\end{figure*}

Figure~\ref{fig:modes_oam_turb} provides a visualization of the effect of different turbulence levels on the input OAM beams, from weak~(a) to moderate~(b) to strong~(c) turbulence. As a representative example, we plot the realization corresponding to the maximum error within the same turbulence strength, i.e., the one associated with the larger beam distortion. It is worth noticing that the statistical contribution of each Zernike mode decreases with the mode index~\cite{Noll:76}. Accordingly, low-rank aberrations such as tip and tilt typically dominate the beam dynamics, resulting in an overall decentering effect. This is also evident in Fig.~\ref{fig:modes_angle_turb}, showing the effect of the same aberrations on ANGLE modes.

\begin{figure*}[!htb]
	\begin{center}
		\includegraphics[width=0.8\linewidth]{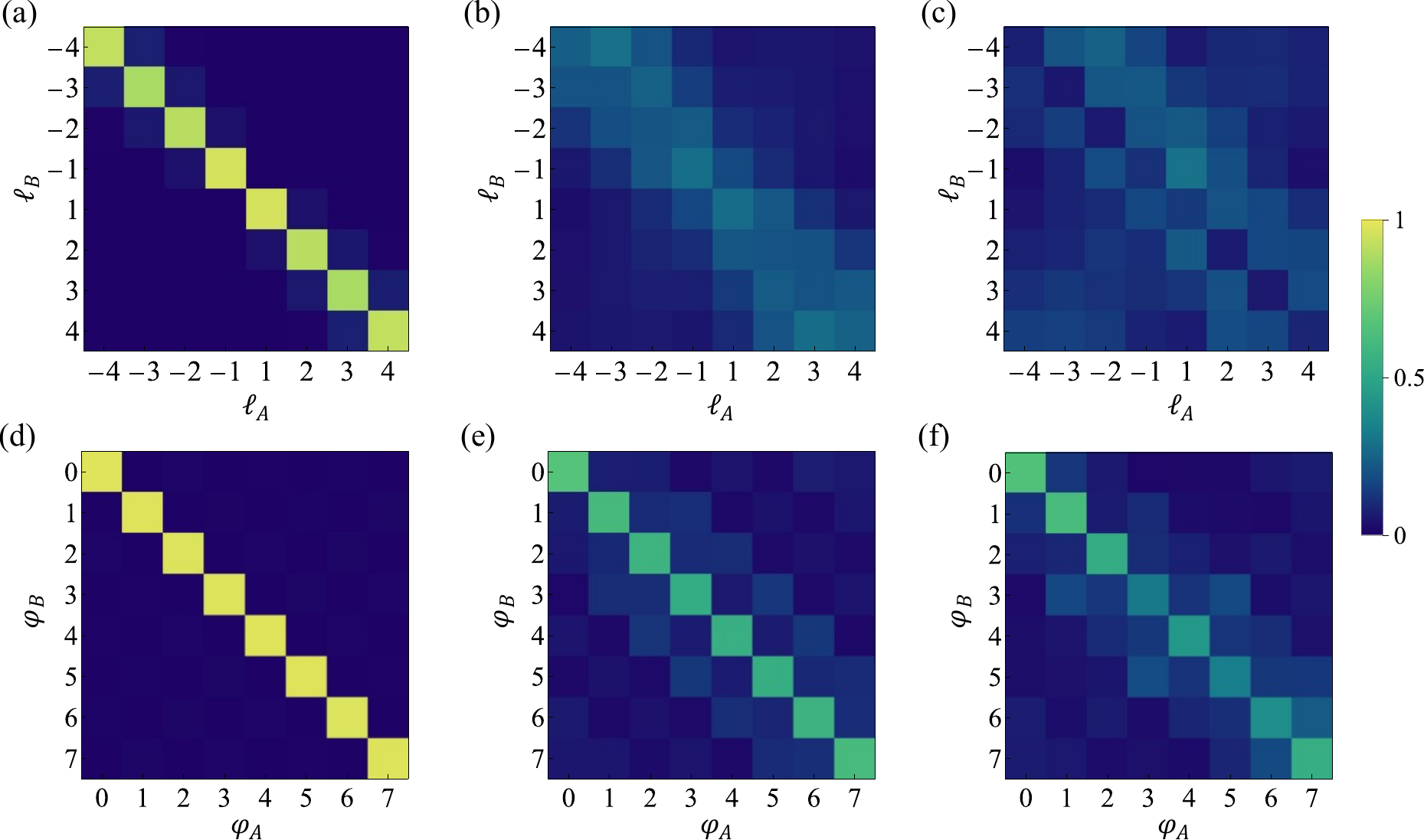}
		\caption{\textbf{Crosstalk matrix.} Average crosstalk integrals between different OAM and ANGLE modes across a turbulent channel. The average is computed over 100 realizations of (a-d)~weak ($C_n^2=10^{-16}$), (b-e)~moderate ($C_n^2=10^{-15}$), and (c-f)~strong turbulence ($C_n^2=10^{-14}$). Each row of individual panels refers to the normalized average crosstalk over 100 independent realizations.}
		\label{fig:crosstalk}
	\end{center}
\end{figure*}

The crosstalk integrals between the modes are plotted in Fig.~\ref{fig:crosstalk}. In the case of weak turbulence (a-d), both bases guarantee secure communications. Interestingly, for moderate (b-e) and strong (c-f) turbulence, the ANGLE basis proves more robust than OAM. This is ascribed to the concentration of optical power in a smaller region, associated with a reduced effective beam waist. More rigorously, the security of a certain basis across a channel is quantified by the Quantum Dit Error Rate (QDER), which is the percentage of incorrect measurements in the shared key after the security verification. In 8 dimensions, a QDER greater than 24.7\% results in the impossibility of guaranteeing secure communications as the key rate (given in Eq.~\eqref{eq:keyrate}) will be negative. This value is related to the amount of information that can be transferred per photon sent through the channel. The average QDERs extracted from the numerical simulation are provided in Table~\ref{tab:QDEROAM}, where the information capacity is expressed in \qo{bits per photon} (b/p). An ideal 8-dimensional channel would support 3~b/p of information capacity. As expected, secure communications can only be established in the weak-turbulence regime ($C_n^2=10^{-16}$).

\begin{table}[h]
\begin{tabular}{ |c|c|c|c|c|c| } 
 \hline
 $C_{n}^2$ & OAM QDER & OAM b/p & ANG QDER  &  ANG b/p \\ 
 \hline
 $10^{-16}$ & $8.18\%$ & $1.72$ & $2.33\%$ & $2.54$ \\ \hline
 $2\times10^{-16}$ & $17.8\%$ & $0.64$ & $5.40\%$ & $2.09$ \\ \hline
 $5\times10^{-16}$ & $55.07\%$ & $0$ & $21.77\%$ & $0.266$ \\ \hline
 $10^{-15}$ & $77.00\%$ & $0$ & $41.47\%$ & $0$ \\ \hline
 $10^{-14}$ & $92.07\%$ & $0$ & $51.96\%$ & $0$ \\ \hline
\end{tabular}
\caption{
Average QDER and information capacity computed for the OAM and ANGLE (ANG) bases from 100 numerical realizations of different turbulence levels over the channel. Security is guaranteed only in the low-turbulence regime. Negative information capacity is reported as 0.}
\label{tab:QDEROAM}
\end{table}

\section{Conclusion}

We employed a recurrent neural network to forecast future turbulence conditions within an optical channel. 
The accuracy of the predictions over significantly long periods demonstrates that our surrogate model has achieved a robust learning of the temporal variation of $C_{n}^{2}$ values correlated to relevant weather parameters. 
Moreover, we have shown how the predictions from TAROCCO can be used to foresee the error rate of a QKD experiment, by simulating a high-dimensional protocol within our free-space link. This result could also apply to QKD ground-to-satellite systems to optimize key exchange rates. While numerical simulations have only been performed for well-known spatial modes of light, future studies could address the performance of other encoding schemes for QKD under turbulence, such as vector beams and the time-frequency domain.

Scintillometer data acquisition will continue over the region of Ottawa, continually expanding the current dataset for enhanced training. Additionally, improved performance could be achieved by adopting an autoencoder architecture, similar to those used in machine translation~\cite{zaytar2016sequence, sutskever2014sequence}. To compensate for limited data, it will also be interesting to explore data augmentation techniques, particularly those involving generative adversarial networks~\cite{iwana2021empirical}.\\

\noindent \textbf{Acknowledgments.}
The authors would like to thank Alicia Sit for valuable discussions and for help purchasing the scintillometer system. The authors would also like to thank Nazanin Dehghan, Alessio D'Errico, and Ashlin Jacob for their help in setting up the scintillometer system. Finally, the authors would like to thank Environment and Climate Change Canada for providing us with the meteorological data. In particular, we appreciate the help of John Richard of Applied Climatology Services. This work was supported by Canada Research Chairs; Canada First Research Excellence Fund (CFREF); National Research Council of Canada High-Throughput and Secure Networks (HTSN) Challenge Program; and the Qeyssat User INvestigation Team (QUINT) Alliance Consortia Quantum grant.

\newpage
\clearpage

\appendix

\section{Permutation Feature Importance}
\label{appendix: grid}


 \begin{figure*}[!htb]
	\begin{center}
	    \includegraphics[width=\linewidth]{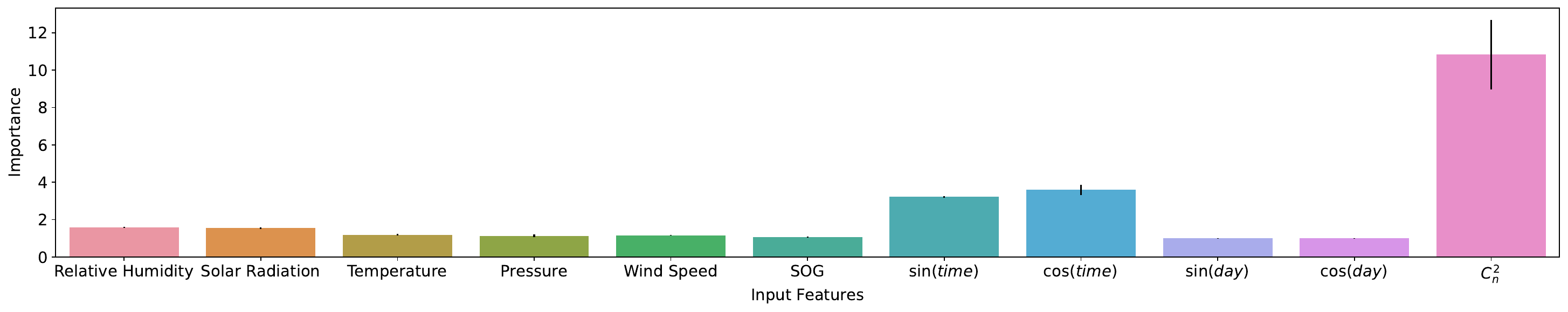}
		\caption{\textbf{Importance of input features.} For each input feature, the mean and standard deviation (error bars) of the importance is computed over three repeated permutations.}
\label{fig:featureImportance}
	\end{center}

\end{figure*}
The importance of each feature is assessed through the Permutation Feature Importance (PFI) technique~\cite{pfiPaper}. Given a model trained with every possible input feature, we corrupt the dataset by permuting one of the features. The importance of each feature is quantified by evaluating the model on the corrupted dataset and comparing its performance with the original model: 
 \begin{equation}
 I=\frac{\epsilon_{\text{perm}}}{\epsilon_{\text{orig}}},
 \end{equation}
where $I$ is the feature importance, and $\epsilon_{\text{perm}}$ and $\epsilon_{\text{orig}}$ are the RMSE of the corrupted and original model, respectively.
In addition to the input features indicated in Fig.~\ref{fig:RNN}, we also considered pressure (hPa), snow on ground~(SOG, cm), wind speed (m/s), and the day according to the Julian Calendar. We then excecute PFI on the first 1000 examples of the dataset. For each feature, we repeat the permutation three times to minimize the effects induced by the variability of the permutation. The results are shown in Fig.~\ref{fig:featureImportance}. Prior values of $C_{n}^{2}$ appear to be the most important feature,  followed by time, relative humidity, and solar radiation. This is in congruence with the analysis carried out in Ref.~\cite{Grose:23}. 
It must be noted that this kind of analysis tends to penalize the importance of correlated features, such as pressure and temperature~\cite{pfiCorrelation}. To address this issue, we permute pressure, temperature, humidity, and wind speed simultaneously, effectively considering their importance together~\cite{correlationTogether}. Indeed, increased values for importance were reported. 



 \begin{figure*}[!htb]
	\begin{center}
	        \includegraphics[width=\linewidth]{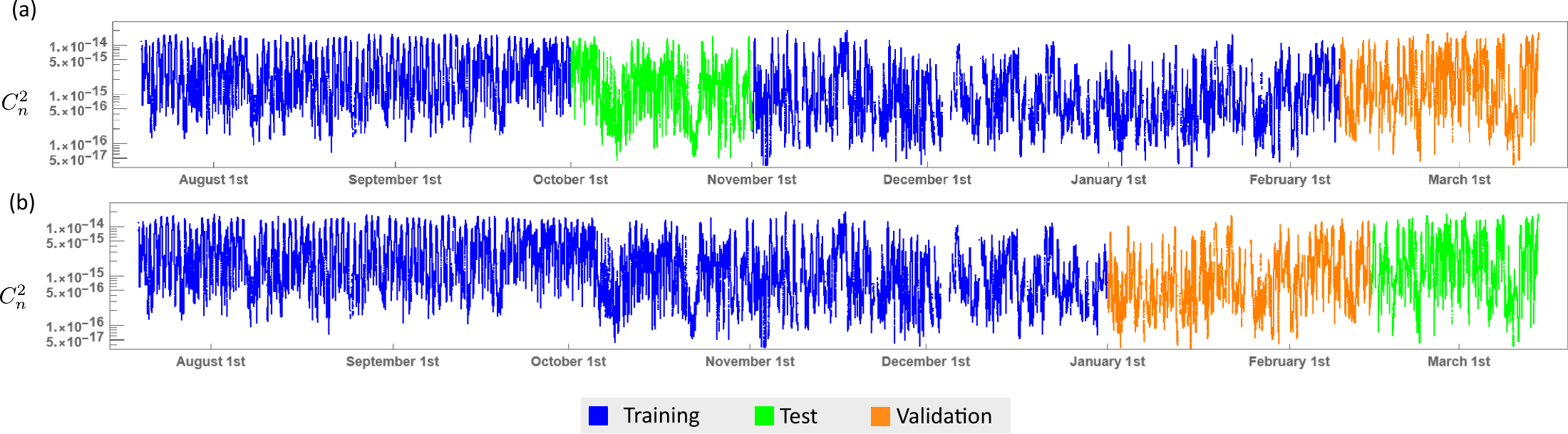}
		\caption{\textbf{Partitions of the complete dataset.} (a) The month of October is selected as test set, and an 85:15 train:validation split is applied to the remaining dataset. (b) A 75:15:10 train:validation:test split is applied to the complete dataset, which corresponds to a test set spanning from mid-February to mid-March (cf.~Sec.~\ref{sec:results}).}
\label{fig:fullData}
	\end{center}

\end{figure*}

\section{Training Hyperparameters}

\label{appendix: train-hype}

\begin{table}[]
\caption{Hyperparameters.}
    \centering
    \renewcommand{\arraystretch}{1.5}
    \begin{tabular}{c | c}
    \hline
\hline
         Hours of Prior Data & 12 Hours  \\
         Forecast Length & 6 Hours \\
         Input Time Resolution & 1 Minute \\
         Output Time Resolution & 15 Minutes \\
         Batch Size & $32$ \\
         Train-Validation Split \textit{(with October)} & 85:15 \\
         Train-Validation-Test Split \textit{(no October)} & 75:15:10 \\
         Initial Learning Rate & $10^{-4}$ \\
         Number of Epochs at Convergence \textit{(with October)} & $149$ \\
         Number of Epochs at Convergence \textit{(no October)} & $262$ \\
         Patience & $15$ Epochs \\ 
         Reduction Factor & $0.1$ \\
         \midrule
         Training Error at Convergence \textit{(with October)} & $1.2 \times 10^{-4}$ \\
         Training Error at Convergence \textit{(no October)} & $7.2 \times 10^{-5}$\\
         Validation Error at Convergence \textit{(with October)} & $1.6 \times 10^{-4}$ \\
         Validation Error at Convergence \textit{(no October)} & $1.3 \times 10^{-4}$ \\
         \bottomrule
         \hline
         \hline
    \end{tabular}
    \label{tab:tab_hyperparams}
\end{table}

The network is trained and validated using data spanning from July 2023 to March 2024. We list the model hyperparameters in Table~\ref{tab:tab_hyperparams}. One month is removed from the original dataset and used for testing (cf.~Sec.~\ref{sec:results}). 
Figure~\ref{fig:fullData} illustrates the partitions of the complete dataset into training, validation, and test datasets, for the two case studies reported in Fig.~\ref{fig:Results}. 

The training process is realized using the Tensorflow library~\cite{tensorflow2015-whitepaper``}, and model optimization is handled using the Adam optimizer~\cite{kingma2017adam}. Training is carried out on the \textit{Narval} supercluster using a NVidia A100SXM4 GPU. An adaptive training strategy is used, whereby the learning rate is reduced by a factor of $0.1$ if the validation loss does not decrease significantly within 15 epochs (this number is referred to as \textit{patience}). The number of hours of prior data is kept fixed at 12. Altogether, training is completed in approximately 5 hours. 
We have also explored longer inputs of 18, 24, and 30 hours. However, no significant reduction in the validation loss was observed. 

We also considered a fully connected neural network, with 2 hidden layers of 1000 neurons separated by ReLU activation layers. Here, the input layer admits a flattened dataset where the first 6 entries represent the input features of the first time step. However, the validation error of our flagship model at convergence is three orders of magnitude smaller than this architecture. 

\section{OAM and ANGLE modes of light}
\label{app:modes}
The 8-dimensional QKD protocol explored in Sec.~\ref{sec:qkd} leverages the MUBs provided by a set of LG modes and the corresponding Fourier-transformed basis. LG modes are labeled by two integers $\ell$ and $p$, representing the azimuthal (OAM) and radial index, respectively. Our protocol only relies on the OAM content of the beam, hence we set ${p=0}$ for all the LG modes. In the position basis, their expression reads
\begin{equation}
\begin{split}    
\braket{x,y,z}{\ell}&=N_\ell\dfrac{w_0}{w(z)}\left(\dfrac{r\sqrt{2}}{w(z)} \right)^\abs{\ell}e^{-r^2/w^2(z)}L_\abs{\ell}^{0}\left(\dfrac{2r^2}{w^2(z)}\right)\\
&e^{-ik\frac{r^2}{2R(z)}}e^{-i\ell\phi}e^{i\psi(z)},
\end{split}
\end{equation}
where $N_\ell$ is a normalization constant, ${r=\sqrt{x^2+y^2}}$ and $\phi=\arctan(y/x)$ are the radial and azimuthal coordinates, respectively, and $L_\ell^{p}$ are the generalized Laguerre polynomials. The beam waist $w(z)$, the radius of curvature $R(z)$ and the Gouy phase $\psi(z)$ can be determined from~\cite{padgett}:
\begin{subequations}
\begin{align}
w(z) &=w_0\sqrt{1+\dfrac{z^2}{z_R^2}} ;\\
R(z) &= z\left(1+\dfrac{z_R^2}{z^2} \right); \\
\psi(z) &= (\abs{\ell}+1)\arctan\dfrac{z}{z_R},
\end{align}
\end{subequations}
where ${z_R=\pi w_0^2/\lambda}$ is the Rayleigh range.

The conjugate basis, referred to as the ANGLE basis, is obtained as
\begin{equation}
\ket{\varphi_j}=\dfrac{1}{\sqrt{d}}\sum_\ell \ket{\ell}e^{2\pi i j g(\ell)/d}, 
\end{equation}
where ${d=8}$, the summations runs from ${\ell=-d/2}$ to ${\ell=d/2}$ excluding the ${\ket{\ell=0}}$ mode, and ${g(\ell)=d/2+(\ell-1)\Theta(\ell)+\ell\Theta(-\ell)}$, with $\Theta$ the Heaviside step function.  


\bibliography{main} 

\newpage
\noindent \textbf{Author Contributions.}
FDC, LS and EK conceived the idea. LS collected and prepared the scintillometer data. TJ developed the neural network architectures. FDC performed numerical simulations of QKD under turbulence. TJ, LS, and FDC wrote the first version of the manuscript. FB, MK, KH, and EK supervised the project.


\end{document}